\documentclass[reprint,superscriptaddress,amsmath,amssymb,aps,pra,showpacs]{revtex4-1}
\usepackage{graphicx,dcolumn,bm}
\usepackage{mcl} 
\usepackage[colorlinks=true]{hyperref} 
\usepackage{cleveref} 

\begin{document}

\title{QED-driven laser absorption}

\author{M. C. Levy}
\email{matthew.levy@physics.ox.ac.uk}
\affiliation{Department of Physics, University of Oxford, Parks Road, Oxford OX1 3PU, UK}
\author{T. G. Blackburn}
\affiliation{Department of Physics, Chalmers University of Technology, SE-41296 Gothenburg, Sweden}
\author{N. Ratan}
\author{J. Sadler}
\affiliation{Department of Physics, University of Oxford, Parks Road, Oxford OX1 3PU, UK}
\author{C. P. Ridgers}
\affiliation{York Plasma Institute, University of York, York, YO10 5DD, UK}
\author{M. Kasim}
\author{L. Ceurvorst}
\author{J. Holloway}
\affiliation{Department of Physics, University of Oxford, Parks Road, Oxford OX1 3PU, UK}
\author{M. G. Baring}
\affiliation{Department of Physics and Astronomy, Rice University, Houston, Texas 77005, USA}
\author{A. R. Bell}
\affiliation{Department of Physics, University of Oxford, Parks Road, Oxford OX1 3PU, UK}
\author{S. H. Glenzer}
\affiliation{SLAC National Accelerator Laboratory, 2575 Sand Hill Road, Menlo Park, CA 94025}
\author{G. Gregori}
\affiliation{Department of Physics, University of Oxford, Parks Road, Oxford OX1 3PU, UK}
\author{A. Ilderton}
\affiliation{Department of Physics, Chalmers University of Technology, SE-41296 Gothenburg, Sweden}
\affiliation{Centre for Mathematical Sciences, Plymouth University, PL4 8AA, UK}
\author{M. Marklund}
\affiliation{Department of Physics, Chalmers University of Technology, SE-41296 Gothenburg, Sweden}
\author{M. Tabak}
\author{S. C. Wilks}
\affiliation{Lawrence Livermore National Laboratory, Livermore, California 94551, USA}

\date{\today}

\begin{abstract}
Absorption covers the physical processes which convert intense photon flux
into energetic particles when a high-power laser illuminates optically-thick matter.
It underpins important petawatt-scale applications today,~\eg  medical-quality
proton beam production. 
However, development of ultra-high-field applications has been hindered since no study so far 
has described absorption throughout the entire transition from the classical to the
quantum electrodynamical (QED) regime of plasma physics. Here we present a model of
absorption that holds over an unprecedented six orders-of-magnitude in optical
intensity and lays the groundwork for QED applications of  laser-driven particle beams.
We demonstrate 58\% efficient \gr~production at \casc ~and the creation of an
anti-matter source achieving $4\times 10^{24}\ \mathrm{positrons}\ \mathrm{cm^{-3}}$,
$10^{6}~\times$ denser than of any known photonic scheme. 
These results will find applications in scaled laboratory probes of black hole and
pulsar winds,
\gr\ radiography for materials science and homeland security, and fundamental
nuclear physics.
\end{abstract}

\maketitle

\tableofcontents

\section{Introduction}

The peak focal intensities produced by high-power laser technology have undergone
rapid increases since the advent of chirped-pulse amplification in the late
1980s~\cite{Strickland1985}. Today, the record intensity has reached up to $10^{22}\ \wcm$
and next-generation facilities such as the Extreme Light Infrastructure (ELI) will
achieve in excess of $10^{24}\ \wcm$ at optical wavelengths~\cite{ELI}. These lasers
have the potential to shed light on the structure of the quantum
vacuum~\cite{Marklund2006,DiPiazza2012}, 
settle an ultimate upper limit to the attainable intensity of an electromagnetic
wave~\cite{Fedotov2010a,Bulanov2010,Grismayer}, discern the transition between classical
and quantum radiation reaction~\cite{DiPiazza2012,Neitz2013,Blackburn2014}, and provide efficient
sources of intense \gr s and dense anti-matter~\cite{Bell2008,Ridgers2012}. 
However, all analyses of optically-thick QED plasma thus far have been restricted to
numerical studies~\cite{Ridgers2012,Benedetti2013,Ji2014b} and simplifed situations
where the ion mass is infinitely-large~\cite{Serebryakov2015}, the laser polarization
is such that the fast-oscillating component of the nonlinear heating force vanishes,
or \gr~ radiation is treated in a classical fashion~\cite{Capdessus2015,Nerush2015,Liseykina}.

By combining strong-field QED with recent advances in plasma kinematic
theory~\cite{Levy2013PoP,Levy2014}, we present the first analytical absorption model
which elucidates how optically-thick matter responds to high-power laser illumination
from the scale of today's petawatt lasers all the way to the QED electron-positron pair
cascade. To span these optical laser intensities this work, for the first time,
self-consistently accounts for quantum mechanically-correct \gr~emission, hydrodynamic
particle injection, and $e^-/e^+$ field-screening  taking place in the partial standing
wave set up by laser absorption.

	\begin{figure*}
	\center{\includegraphics[width=0.8\linewidth]{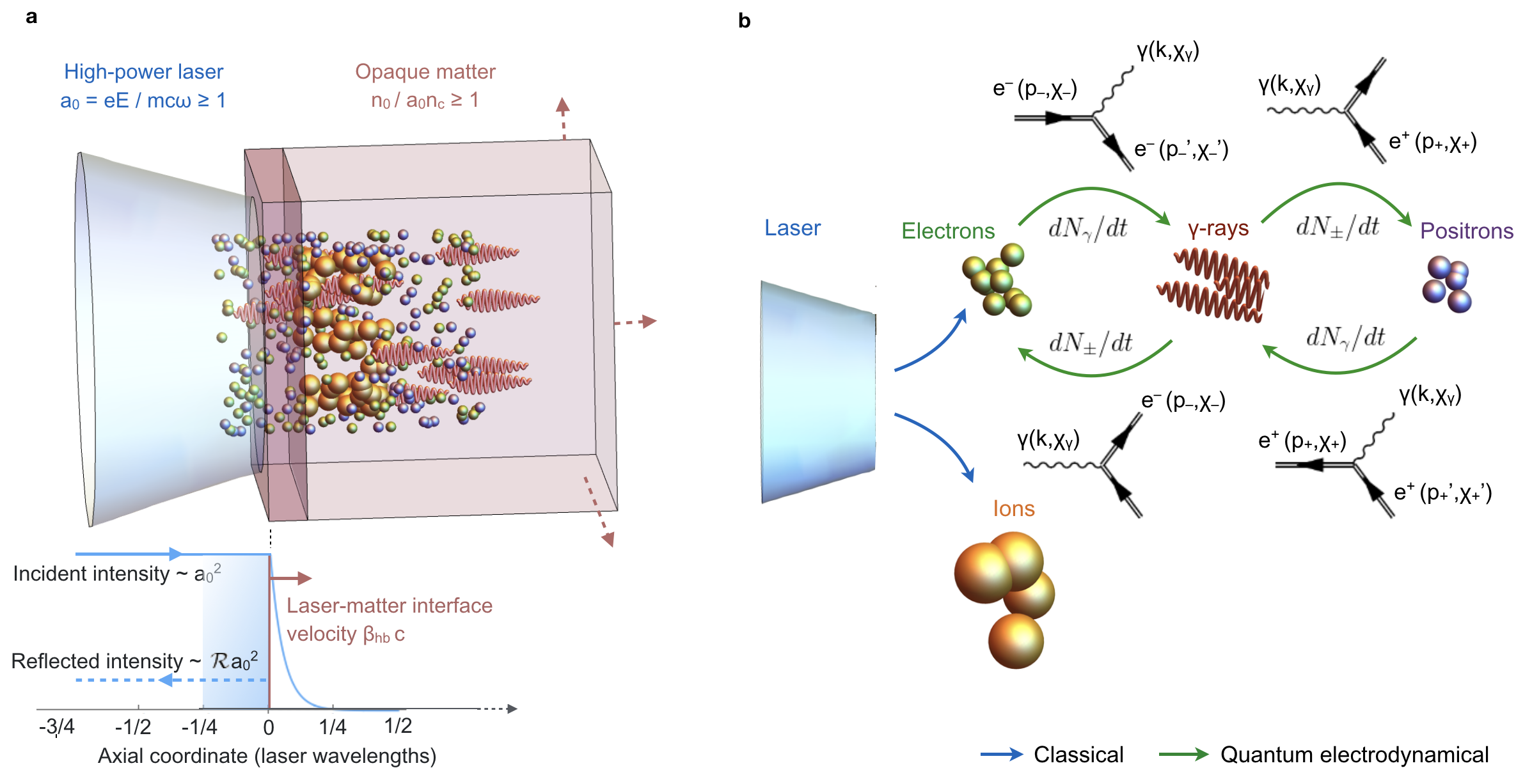}}
	\caption[The response of optically thick matter to high-power laser light]{
		\textbf{The response of optically thick matter to high-power laser light.}
		(a) Relativistic particle flows are produced through absorption when a high-power
		laser is shined onto optically-thick matter. Along the laser-matter interface
		ultrafast field-ionization creates a thin layer of opaque plasma.
		(b) The electrons and ions are accelerated via classical electromagnetic
		absorption processes~\cite{Kemp2012,Levy2014}. 
		Leptons are cooled through the emission of quantum synchrotron \gr~ photons
		in the ultra-strong optical field. The \gr s ``decay'' into electron-positron
		pairs 
		which become electromagnetically accelerated, thereby linking together the
		classical and quantum electrodynamical absorption. Thick double-lines in the
		Feynman diagrams indicate ``dressed'' charged-particle states which account
		to all orders in $a_0$ for the effect of the classical laser field.
		}
	\label{fig:schem}
	\end{figure*}

Key features of the model are depicted schematically in Fig. \ref{fig:schem} (a).
Here a thick slab of opaque matter is illuminated at normal incidence by a linearly-polarized
high-power laser beam having peak 
dimensionless strength parameter $a_0 = e E/m c \omega $, where $E$ is the laser electric
field and $\omega$ is its angular frequency. Over the Lorentz-transformed collisionless skin
depth $\lp$ shown in dark red, the laser rapidly field-ionizes the matter to create a
supercritical plasma, \ie with electron density $\ne$ exceeding the relativistically-correct
critical density $a_0 \ncr$ where $\ncr = \varepsilon_0 m \omega^2/e^2$ ($\varepsilon_0$
is the vacuum permittivity). We use $1\um$ wavelength light and a model geometry centered
at the supercritical interface determined by $\ne / a_0 \ncr \simeq 1$  so that the laser
evanescences over $\lp$ in the downstream matter. $\ne$ is thusly scaled linearly with
$a_0$ to maintain the optically-thick condition, which in the classical regime introduces
a similarity to the electron dynamics~\cite{Pukhov2004}.

The interface itself is accelerated to an appreciable fraction of light-speed $\betaHB$
along the wave propagation vector by pressure associated with the intense photon flux.
This ``hole-boring'' electrostatic process accelerates ions to moderately-relativistic
energies and reduces through hydrodynamic motion the optical photon flux deposited onto
the interface~\cite{Levy2013PoP}. 
In the upstream vacuum region, indicated using blue shading, collisionless classical
and QED absorption processes create \gr s and $e^+$ and accelerate highly-relativistic
$e^-/e^+$ through the interface into the dense matter bulk.

The relevant interactions are therefore highly localized (we show that
the axial lengthscale is smaller than $\laml$) and the instantaneous fractional
absorption into each particle species is equivalent to~\cite{Levy2013PoP,Levy2014},
	\begin{align}
	f_k &= \frac{1}{1-\betaHB}\frac{v_k}{c}\frac{n_k ~ \eavg}{ u_\mathrm{l}},
	&
	f &= \sum_k f_k =  1-\refl,
	\label{eqn:absdef}
	\end{align} 
where $k$ indexes the set of particles shown in Fig. 1 (b): $f_k$ is the absorption,
$n_k$ the number density, $\eavg$  the ensemble-averaged energy, and $v_k$ the
velocity associated with particles of the $k^{th}$ type. $f$ is the total absorption,
$\refl$ is the reflectivity, and $u_\mathrm{l}=a_0^2 m c^2 \ncr/2$ is the energy density
of the laser.

The coupling between strong-field QED and laser-produced dense plasma is illustrated
in Fig. \ref{fig:schem} (b). The classical portion of the interaction involves deposition
of laser energy into electrons and ions as is shown using blue arrows. QED processes,
indicated using green arrows, redistribute energy coupled into relativistic electrons
first through the emission of quantum synchrotron \gr s by propagation in the laser field.
QED corrections to the electron absorption are thus introduced as the Lorentz-invariant
quantum parameter,
	\begin{equation} 
	\chi = \frac{\gamma}{\Esch} \sqrt{(\mathbf{E} + \mathbf{v}\times\mathbf{B})^2 - (\mathbf{E}\cdot\mathbf{v}/c)^2 },
	\label{eqn:eta1}
	\end{equation} 
reaches $\sim 0.1$~ ($\mathbf{B}$ is the laser's magnetic field,
$\Esch = 1.3\times 10^{18}~\mathrm{V\ m^{-1}}$ is the critical field of
QED~\cite{Heisenberg1936}. 
Production of electron-positron pairs takes place when $\chi \sim 1$ through \gr~conversion
in the laser fields themselves which is controlled by the quantum parameter
$\chi_\gamma = \hbar \omega/(2 m c^2  \ESch) \ \left| \mathbf{E}_\perp +
\hat{\mathbf{k}}\times c\mathbf{B} \right|$, where the photon has energy $\hbar \omega$
and propagates along the $\hat{\mathbf{k}}$ unit vector, and $\mathbf{E}_\perp$ is
the component of the laser electric field normal to its motion. The \gr~absorption
is thus set by the leptonic radiative energy loss and the positrons take an $n_+/\nh$
fraction of the total pair energy which is always less than $\frac{1}{2}$.

\section{Absorption Model}

\subsection{Classical Relativistic Absorption}

At the petawatt scale, the theoretical extrema of $f$ in the supercritical situation
have recently been reported using a kinematic interaction model. This works by applying
shockwave-like conservation laws across the supercritical interface to constrain $f$
in relation to the laser and unperturbed matter properties~\cite{Levy2013PoP,Levy2014}.
In this section we review the essential results and extend them to the situation of
interest here.

Denoting the absence of QED using a superscript $0$, the sets of all possible
$f$-curves associated with \cref{eqn:absdef} are given by
	\begin{align}
	\begin{split}
	f_i^0 &= \frac{2 \bpz \refl^{3/2}}{\sqrt{\bpz^2 \refl+1}-\bpz \sqrt{\refl}}
	\\
	f_e^0 &= \frac{(1-\refl) \sqrt{\bpz^2 \refl+1} - (1+\refl) \bpz \sqrt{\refl}}{\sqrt{\bpz^2 \refl+1}-\bpz \sqrt{\refl}}
				+ \mathop{O}\!\left( \frac{\rh}{\bpz^2} \right).
	\end{split}
	\end{align}
Here $f^0 = f_e^0 + f_i^0$ where the terms correspond to electrons and ions, respectively.
These limits bound the efficiency of nonlinear absorption mechanisms operating in
petawatt-scale laser interactions with supercritical matter~\cite{Haines2009,Kemp2009,%
May2011,Kemp2013}, while effects occurring downstream of the interface such as efficient
plasma wave heating of dense matter~\cite{Lefebvre1997, Sherlock2013Heating} are
abstracted from the analysis.

As a consequence, the nonlinear plasma dynamical interaction is parameterized by three
control variables: $\rh = \nh/\ni \ m /  (M+Z m) \ll 1$, corresponding to the
normalized mass density of relativistic electrons accelerated by the laser, the
reflectivity $\refl$, and the dimensionless velocity $\bpz=\left[Z m \ncr/(2 M \ne)\right]^{1/2} a_0$,
and the supercritical interface has uniform charge-state $Z$, and ion density
$\ni = Z \ne$ and mass $M$. 
We note that $a_0$ corresponds to the peak laser dimensionless strength parameter,
so the laser-cycle-averaged value often used in classical laser interactions is recovered
by $a_0 \rightarrow \sqrt{2}a_0$.
The dimensionless velocity of the supercritical interface is given by~\cite{Levy2013PoP,Levy2014}  
	\begin{equation}
	\betaHB = \sqrt{\frac{\refl \bpz^2}{1+\refl \bpz^2}}.
	\label{eqn:hbVel}
	\end{equation}

The laser temporal envelope is constrained by
$\il\ (\partial \il/\partial t)^{-1} \gg 2\pi \ \omega_{\mathrm{pe}}^{-1}$ which is readily
achieved in realistic conditions, where the electron plasma frequency
$\omega_{\mathrm{pe}}= (\ne e^2  / \varepsilon_0 m)^{1/2}$.
Damping of transient momentum effects also requires that the laser pulse duration
$\tau_\mathrm{l}$ satisfies $\tau_\mathrm{l}\ \omega_{\mathrm{pi}} > 2 \pi C$ where
the ion plasma frequency $\omega_{\mathrm{pi}}= (Z \ne e^2  / \varepsilon_0 M)^{1/2}$
for supercritical interface uniform charge-state $Z$,  ion mass $M$, and $C\simeq 3-5$.
The target thickness \mcl{$D$} should exceed the hole-boring depth and the effective
relativistic electron refluxing range,
$D > c[ \tau_\mathrm{l}/2 + \int_0^{\tau_\mathrm{l}} \betaHB\ dt ]$ in order to
access the steady-state interaction~\cite{Levy2013PoP,Levy2014}.
The rate of electron heating in the model is proportional to the scalar product
of the particle velocity and electric field which scales linearly with $a_0$. 
In the absence of radiative cooling this implies ponderomotive scaling of the electron
energy~\cite{Wilks1992,May2011}, \ie $\langle \mathcal{E}_e^0 \rangle = a_0 m c^2
/\sqrt{2}$,
which Fig.~\ref{fig:temp} shows is consistent with the energetics exhibited in our
simulations.

To first order in $\rh$, the absorption is then given by,
	\begin{align}
	f_e^0 &\simeq \frac{1+\bpz}{\bpz^2} \left(\frac{3\sqrt{2}}{20} a_0 -1 \right) \rh,
	\label{eqn:feClass2}
	\\
	f_i^0 &\simeq \frac{2 \bpz}{1+2\bpz}
			- \frac{4\bpz^2 + 7\bpz + 3}{2\bpz (1+2\bpz)^2} \left(\frac{3 \sqrt{2}}{10} a_0-1\right) \rh,
	\label{eqn:fiClass2}
	\end{align}
where the (laboratory frame) reflectivity $\refl = 1-f_e^0-f_i^0$.
When $a_0 \rightarrow 10 \sqrt{2}/3$ the model suggests that absorption becomes
asymptotically small. This implies that for $a_0 \lesssim 5$ in the
supercritical situation the electron dynamics can be non-ponderomotive,
as has been suggested in previous works~\cite{Beg1997,Haines2009,Kluge2011}.
We consider higher intensities in the present work which means that the supercritical
plasma comprising the interface reaches a fully-ionized state and
\cref{eqn:feClass2,eqn:fiClass2} are controlled by $a_0$ and the
$\rh \propto \nh/\ne$ parameter alone. Interactions between $a_0<1$ laser light
and opaque matter involve collisional dynamics~\cite{Wilks1997} and are outside the
scope of this work, as the threshold for a high-power laser interaction is set by
$a_0 \geq 1$ when the oscillatory velocity of a free electron approaches $c$.

\subsection{Strong-Field QED Processes}

	\begin{figure}
	\centering
	\includegraphics[width=0.8\linewidth]{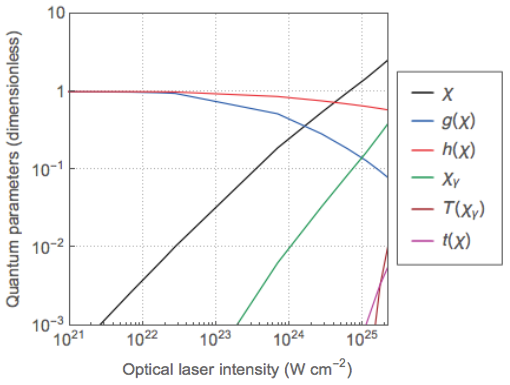}
	\caption[Quantum parameters controlling particle production]{
		\textbf{Quantum parameters controlling particle production.}
		Quantum corrections to absorption become important when the
		Lorentz-invariant parameters $\chi$ and $\chi_\gamma \gtrsim 0.1$.
		$\chi$ compares the electric field strength in the rest
		frame of a lepton to the quantum critical field~\cite{Heisenberg1936}. 
		Leptons emit $N_\gamma \propto \chi h(\chi)$ photons per
		unit time and lose energy at a rate of $\Power \propto \chi^2
		g(\chi)$. $h(\chi)$ and $g(\chi)$ thus describe quantum
		corrections to the photon spectrum which affect the classical
		emissivities. $\chi_\gamma$ controls the probability for a
		\gr~to convert into an electron-positron pair by direct
		interaction with the laser fields. Pair creation is suppressed
		for $\chi_\gamma \lesssim 0.1$, but becomes exponentially
		more probable as $\chi_\gamma$ increases, leading to a ``cascade''
		of self-created particles when $\chi_\gamma \gtrsim 1$.
		Its exponential growth is parameterized by the auxiliary
		function $T(\chi_\gamma)$ in the pair creation rate, and
		$t(\chi)$ approximates $T(\chi_\gamma)$ with $\chi_\gamma$
		weighted over the emission spectrum (details provided in the text).
		}
	\label{fig:qed}
	\end{figure}

Inspection of \cref{eqn:eta1} reveals the crucial role played by the reflected
laser wave, since a single laser pulse will accelerate electrons from rest such
that they co-propagate along the direction of the light wavevector.  
In this situation the electric and magnetic forces acting on the electron
approximately cancel one another -- which leaves $\chi \ll 1$ for all forseeable
laser systems~\cite{Bell2008}.

However, this picture changes substantially when the partially-reflected wave
is treated. We work in the frame of reference co-propagating with the supercritical
interface at $\betaHB c$, meaning that the supercritical matter is injected into
the laser fields with axial velocity $-\betaHB c$, the surface has reflectivity
$\RHB = \refl (1+\betaHB)/(1-\betaHB)$, the incident and reflected waves have
frequency $\omegaHB = \omega [(1+\betaHB)/(1-\betaHB)]^{-1/2}$. At focus the
laser spot size is much larger than the wavelength $\laml$ and we assume the laser
ponderomotive force rapidly sweeps away any underdense pre-plasma associated with
the production of the laser pulse for the ultra-high intensities of interest
here~\cite{Macchi2013a}. The electric and magnetic fields in the vacuum region are
given by,
	\begin{align}
	\begin{split}
		\frac{e E_x}{m c \omegaHB} &=
		a_0 \left[\cos(\xi - \tau) -
		\sqrt{\RHB} \cos (\xi + \tau)\right],
	\\
	\frac{e B_y}{m \omegaHB} &=
		a_0 \left[\cos(\xi - \tau) +
		\sqrt{\RHB} \cos (\xi + \tau)\right],
	\end{split}
	\label{eq:VacuumFields}
	\end{align}
defining the dimensionless axial distance $\xi= \omegaHB z / c$ and time
$\tau = \omegaHB t$.

$\nh$ electrons escape the bulk plasma and enter the vacuum region where
the partial standing wave can accelerate them to relativistic energy,
radiating \gr s if  $a_0$ is large enough. For $\RHB = 1$, $z = -\laml/4$
is a node of the magnetic field and an antinode of the electric field.
The electron is accelerated parallel to the electric field, so $\chi$
vanishes and it reaches a peak momentum of $\sim 2 a_0 m c$~\cite{Kemp2009,May2011}.
However, the $B = 0$ node is unstable -- if the electron starts
slightly off-node, it is pushed towards the electric field node.
This suggests only particles within the region $-\laml/4 \lesssim z \lesssim \lp$
are heated by the wave, meaning that absorption is highly localized with
an axial lengthscale less than $\laml$.

We assume the electron is ultrarelativistic with $v = c$, and that the
electric and magnetic forces perpendicular to its motion are almost in balance.
The particle's relativistic motion means its period of oscillation is much
longer than the laser period. Furthermore, due to the localization,
the wave phase interval in which it is accelerated is very small. As a
consequence of this, averaged over the wave phase, we can anticipate a
steady-state solution for the absorption. Confining the electron motion to
the $x$-$z$ plane, force balance in the perpendicular direction gives
$E_x \cos \theta \simeq c B_y$ where $\theta$ corresponds to the angle
of the electron's acceleration relative to the laser axis. This defines
the trajectory by $d\xi/d\tau = c B_y / E_x$, with boundary condition
that $\xi = -\pi/2$ at $\tau = \pi$. Substituting $\xi = -\pi/2
+ \cos\theta (\tau - \pi)$ into the field definitions
given by equation (\ref{eq:VacuumFields}) and expanding around $\tau = \pi$, we
find that the electrons are initially accelerated at an
angle $\cos\theta = (1-\RHB^{1/4})/(1+\RHB^{1/4})$. This angle cannot
be sustained indefinitely -- the magnetic field increases along this
trajectory and therefore the electron should be deflected toward the
supercritical interface. We make the further approximation that the
electron propagates at this angle until such a time that $E_x (\tau,
\xi) = c B_y (\tau, \xi)$. Thereafter it moves parallel to
the laser axis until it crosses the interface at $\xi = 0$.

In reality, the complex nature of the fields and the stochastic
nature of photon emission mean that the electron will deviate
from the balanced trajectory -- nevertheless, this rectilinear model should
suffice to describe the most important physics.

\subsubsection{Quantum \gr~Radiation}

The instantaneous average power radiated to photons by an electron (or positron) is 
$\Power = (2\alpha/3\tC) \chi^2 g(\chi) mc^2$, where $\alpha\simeq 1/137$ is the fine
structure constant, $\tC=\hbar/m c^2$ is the Compton time and the auxiliary
function $g(\chi)$ is the factor by which quantum corrections reduce the radiated
power~\cite{Ritus1985,Ridgers2014}. Although the fact that photon emission is a stochastic
process means $\chi(\tau)$ does not change smoothly over the interval
$\tau \in [\pi, \tau(\xi=0)]$, it is possible to use the average power to determine
the average radiated energy~\cite{Ridgers2014}, by $\langle \Energy \rangle = \int
\Power (\langle\chi\rangle)\,\rmd \tau$. 

Assuming for the present $\langle \mathcal{E}_e^0 \rangle = a_0 m c^2/\sqrt{2}$
is constant over the electron motion, the quantum parameter is given by
$
	\chi(\tau) =
		a_0 |c B_y - E_x \cos\theta|/(\sqrt{2} \ESch)
$.
Squaring and integrating this along the electron trajectory defines the
root-mean-square (r.m.s.) $\chi_\text{rms}$ by
$\int \chi^2 \rmd \tau \equiv \chi_\text{rms}^2 \Delta \tau$, yielding,
	\begin{widetext}
	\begin{equation}
	\chi_\text{rms} =
		\frac{a_0^2}{\pi}
		\frac{\hbar \omega \sqrt{\frac{1-\betaHB}{1+\betaHB}}}{m c^2}
		\left\{
			\frac{2r^2[\pi(1+r^3-r^5-r^6)-4r\sin(\pi r)]}{(1-r)(1+r)^2(1+2r)^2}
			- \frac{r(1+r^3+r^4)\sin(2\pi r)}{(1+r)(1+2r)^2}
		\right\}^{1/2},
	\label{eq:rmsEta}
	\end{equation}
	\end{widetext}
where $r^4 \equiv \RHB$ \mclb{and $\omega$ is the laser angular frequency
in the laboratory frame.} As can be seen in Fig.~\ref{fig:qed}, $\chi$ is sufficiently
large that $g(\chi)$ gives a non-negligible correction to the
radiated intensity. We include this by using $\chi_\text{rms}$ to define an
average $\langle g\rangle \equiv g(\chi_\text{rms})$ with which we scale
the radiated energy.

\begin{figure}
	\centering
	\includegraphics[width=0.8\linewidth]{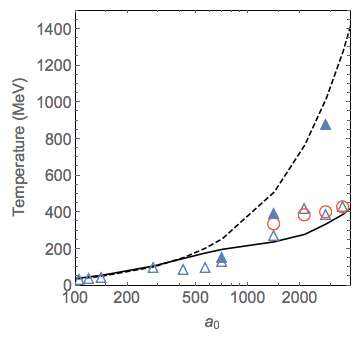}
	\caption[Electron and positron temperature scalings]{
		\textbf{Electron and positron temperature scalings.}
		Simulation results are shown for electrons using blue markers and positrons
		using red markers.  The filled symbols correspond to simulations in which
		the QED module was deactivated and the open symbols to simulations in
		which the QED module was activated. The dashed black curve corresponds
		to the ponderomotive temperature scaling~\cite{Wilks1992} and the solid
		black curve to the cooled lepton temperature scaling given by
		\cref{eq:RadiatedEnergyWithRecoil}.
		}
	\label{fig:temp}
	\end{figure}

To account for the recoil of the electron, we consider the evolution
equation $\gamma'(\tau) = -A \gamma^2$, since $\chi \propto \gamma$.
Assuming $A$ to be constant and defining $\gamma_0\equiv \gamma(\tau=0)$
means that $\gamma(\tau) = \gamma_0 / (1 + A \gamma_0 \tau)$. We can identify
$A \gamma_0^2 \tau$ with the radiated energy in absence of recoil
$\langle\Energy_{\gamma,\text{hb}}\rangle$ andtherefore calculate the 
recoil-corrected quantities as
	\begin{align}
	\chi^\text{recoil}_\text{rms} &=
	\frac{\chi_\text{rms}}{ \sqrt{1 + \langle\Energy_{\gamma,\text{hb}}\rangle/\langle \Energy_\text{e}^0 \rangle}},
	\label{eq:RadiatedEnergyWithRecoil}
	\\
	\langle \Energy_\pm \rangle &=
	\frac{\sqrt{ \frac{1+\betaHB}{1-\betaHB} } \langle \Energy_\text{e}^0 \rangle}{1 + \langle\Energy_{\gamma,\text{hb}}\rangle/\langle \Energy_\text{e}^0 \rangle}.
	\label{eqn:etaRmsRecoil}
	\end{align}
$\langle \Energy_\pm \rangle$ corresponds to the radiatively-cooled $e^-/e^+$ energy
and the Doppler factor accounts for the transformation
from the co-propagating to laboratory frame;
the radiated energy in the absence of recoil is given by,
	\begin{equation}
	\langle\Energy_{\gamma,\text{hb}}\rangle =
		\frac{\pi\alpha}{3}
		(1 + 2 r)
		\chi_\text{rms}^2 \langle g \rangle
		\frac{(mc^2)^2}{\hbar\omega \sqrt{\frac{1-\betaHB}{1+\betaHB}}}.
		\label{eqn:erad}
	\end{equation}
We compare this expression for the temperature with the ponderomotive
prediction and with simulation data in Fig. \ref{fig:temp}.

\subsubsection{Electron-Positron Pair Production}
\label{subsubsec:PairProduction}

The differential probability rate for a \gr~photon, travelling at an angle
$\theta$ to the laser axis, to convert to an electron-positron pair via
the multi-photon Breit-Wheeler process~\cite{Erber1966,BaierKatkovStrakhovenko}
is given by
	\begin{equation}
	\frac{dP_\pm}{d t} =
		\frac{\alpha}{\tC} \frac{mc^2}{\hbar \omega}{\chi_\gamma}
		\, T (\chi_\gamma)
	\label{eq:PhotonDecayRate}
	\end{equation}
$T(\chi_\gamma)$ is the Breit-Wheeler auxiliary function which is well
approximated as~\cite{Erber1966,Kirk2009},
	\begin{equation}
	T (\chi_\gamma) \simeq
	\frac{0.16}{\chi_\gamma}K^2_{1/3}\left( \frac{2}{3\chi_\gamma} \right)
	\label{eq:bwaux}
	\end{equation}
where $K$ is the modified Bessel function of the second kind.
On the other hand, second-order pair production via virtual photons in
the Trident process is strongly suppressed in the laser-plasma context
and is negligible for our purposes~\cite{King2013}.
$T(\chi_\gamma)$ in equation (\ref{eq:PhotonDecayRate}) should be
weighted using the full synchrotron spectrum, \ie using the transformation,
	\begin{equation}
	T(\chi_\gamma) \rightarrow
	\int_0^{\chi/2}\!
		\frac{F(\chi,\chi_\gamma)\,T (\chi_\gamma)}{h(\chi)\,\chi_\gamma}
	\,\rmd\chi_\gamma
	\label{eqn:tchi}
	\end{equation}
so that the photon spectrum is characterized by $\chi$ of the radiating electron.
The integral in equation (\ref{eqn:tchi}) in general must be evaluated
numerically, but over the range $0 \leq \chi \leq 10$ the following
analytical approximation differs from the true value by at most
$5\%$:
	\begin{equation}
	t(\chi) \simeq \frac{3}{16} \exp \left(
		-\frac{114}{37 \sqrt{\chi}}
		- \frac{71}{18 \chi} - \frac{1}{19 \chi^2}
		\right)	
	\label{eqn:tchi2}
	\end{equation}
The intensity scalings of $\chi_\gamma$ and the auxiliary functions
\cref{eqn:tchi,eqn:tchi2} are shown in Fig.~\ref{fig:qed}.

The average angle between \gr~wavevectorand the laser axis is given by 
$\cos\langle\theta_{\gamma,\text{hb}}\rangle \simeq \sqrt{3}(1-\RHB)/2$.
Together with the fact that $\chi^\text{recoil}_\text{rms}$ characterizes
the shape of the photon spectrum, we can integrate the photon conversion
rate along the trajectory defined by this angle to obtain the pair
creation probability
	\begin{equation}
	P_\pm \simeq
		\frac{\alpha a_0 \RHB^{3/4} \left( 1 - \RHB + 2/\sqrt{3} \right) t(\chi^\text{recoil}_\text{rms})}
			{\left(1-\RHB\right)\left(1+\RHB^{1/4}\right)}.
	\label{eq:PairProbability}
	\end{equation}
For \gr~ number density $n_\gamma$, equation (\ref{eq:PairProbability})
controls the positron density according to 
$n_+/\ne = P_\pm n_\gamma / \ne$. $n_\gamma$ is obtained
by integrating the emission rate along the electron trajectory and then
scaling by the number of electrons that have escaped the bulk matter:
	\begin{equation}
	\frac{n_\gamma}{\ne} =
		\frac{\nh}{\ne}
		\frac{5\alpha}{2\sqrt{3}} a_0
		\left[
			1 - r^2 - \left(1 + r^2\right)
			\cos \left(\pi r\right)
		\right]
		\frac{h(\chi^\text{recoil}_\text{rms})}{5\pi/3}.
	\label{eq:PhotonYield}
	\end{equation}
	
	\begin{figure*}
	\center{\includegraphics[width=0.8\linewidth]{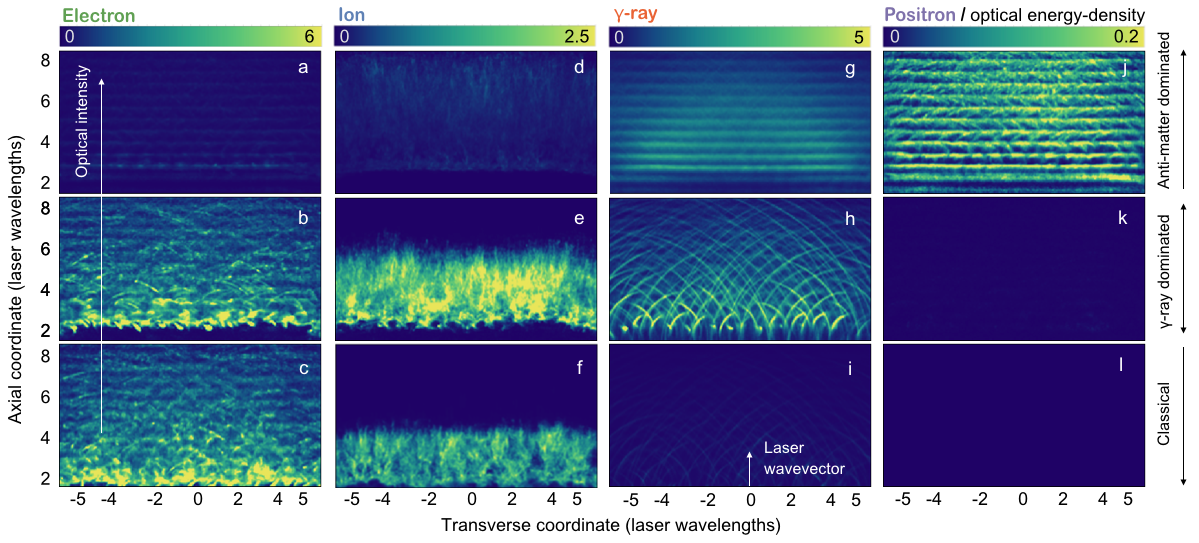}}
	\caption[Particle production]{
		\textbf{Particle production.}
		Snapshots are shown of the normalized spatial energy-density of electrons,
		ions, \gr s, and positrons. 
		Regimes of particle production are illustrated by varying the optical
		intensity from $6.5\times 10^{23}~\wcm$, to $2.6\times 10^{24}~\wcm$,
		and $1.7\times 10^{25}~\wcm$ (rows, bottom to top).
		Thin, high energy-density radial filaments of \gr s are created (h)
		through quantum synchrotron cooling of electrons (b) which penetrate
		the surface plasma layer into the ultra-strong laser field twice each
		optical cycle. Positron production through \gr ~ decay 
		(j-l) and collimation of \gr ~ emission  along the laser wavevector
		increase with intensity.  As a consequence, the normalized
		\gr ~ energy-density peaks around (h) then decreases with intensity, 
		as is shown in (g). All parameters are detailed in the text.
		}
	\label{fig:grb}
	\end{figure*}
\subsection{Feedback Between QED and Classical Dynamics}

QED processes have the effect of redistributing energy initially coupled by
the laser into relativistic electrons. Normalized to the classical electron
value, the \gr~absorption is set by the radiative energy loss of $e^\pm$ as 
    $f_\gamma = \left( 1-\langle \mathcal{E}_\pm \rangle/\epond  \right)\ f_e^0  $.
The complement of this corresponds to the total $e^\pm$ absorption accounting
for \gr~emission as~
    $f_e + f_+ = \langle \mathcal{E}_\pm \rangle / \epond\ f_e^0$.
Positrons take an $n_+/\nh$ fraction of the total pair energy which is always
less than $\frac{1}{2}$ as~
    $f_+ = P_\pm\ (n_\gamma/\nh) \ (\langle \mathcal{E}_\pm \rangle/\epond)\ f_e^0$.
    
Fig.~\ref{fig:temp} shows that electrons in the no-QED simulations are
accelerated to approximately the ponderomotive potential~\cite{Wilks1992},
in good agreement with the classical absorption model. With QED effects present,
the highest energy \gr s are converted by the laser fields into ``daughter''
electrons and positrons which each gain approximately half the parent lepton's energy.
The daughter particles are re-accelerated to $\langle \mathcal{E}_e^0 \rangle$
and cool to $\langle \mathcal{E}_\pm \rangle$ in steady-state.
There is also close agreement between the QED simulation results
and the model predictions for $\langle \mathcal{E}_\pm \rangle$ given by
\cref{eq:RadiatedEnergyWithRecoil} for both positrons and electrons.

Equilibration with electrons extracted from the target can be understood
by considering the work done on the self-created pairs by the laser field through  
$d \mathcal{E} /dt = e \mathbf{v} \cdot \mathbf{E}$. We assume for simplicity
acceleration parallel to the laser electric field in the hole-boring frame
and energy $\mathcal{E} \gg m_e c^2$. The timescale $\tau_{\text{accel}}$
for acceleration to the ponderomotive energy can then be estimated as
$\langle \mathcal{E}_e^0 \rangle /(2 \tau_{\text{accel}}) \sim m_e c^2 a_0 \omegaHB$,
which yields $\tau_{\text{accel}} \sim 1/(2 \sqrt{2} \omegaHB)$.

The timescale associated with radiative cooling~\cite{Fedotov2010a}, $\tau_\gamma$,
is proportional to $1/W_{e \rightarrow \gamma}$ where
$d W_{e \rightarrow \gamma}/d\chi_\gamma = \sqrt{3} \alpha \chi F(\chi, \chi_\gamma)/(2 \pi \tau_c \gamma \chi_\gamma)$
corresponds to the differential rate of photon emission~\cite{Baier1968a,Erber1966,Fedotov2010a}.
Integrating over all photon energies in the range $0\leq \chi_\gamma\leq \chi/2$
using $\gamma = a_0/\sqrt{2}$ gives the total emission rate to be
$W_{e \rightarrow \gamma} = \sqrt{3/2} \alpha \chi h(\chi) / (\pi a_0 \tau_c)$. 
Using $\chi$ and $h(\chi)$ as given in Fig.~\ref{fig:qed},
$\tau_{\text{accel}} \lesssim \tau_\gamma$ for all values of $a_0$ below the pair
cascade. 
A characteristic of S (``shower'') type pair cascades is that daughter positrons
are much cooler than the parent particles due to energy partitioning. By contrast,
these results indicate that electrons and positrons reach comparable energies,
providing a signature of re-acceleration and hence an A (``avalanche'') type
cascade process~\cite{Mironov}.

	\begin{figure*}
	\center{\includegraphics[width=0.8\linewidth]{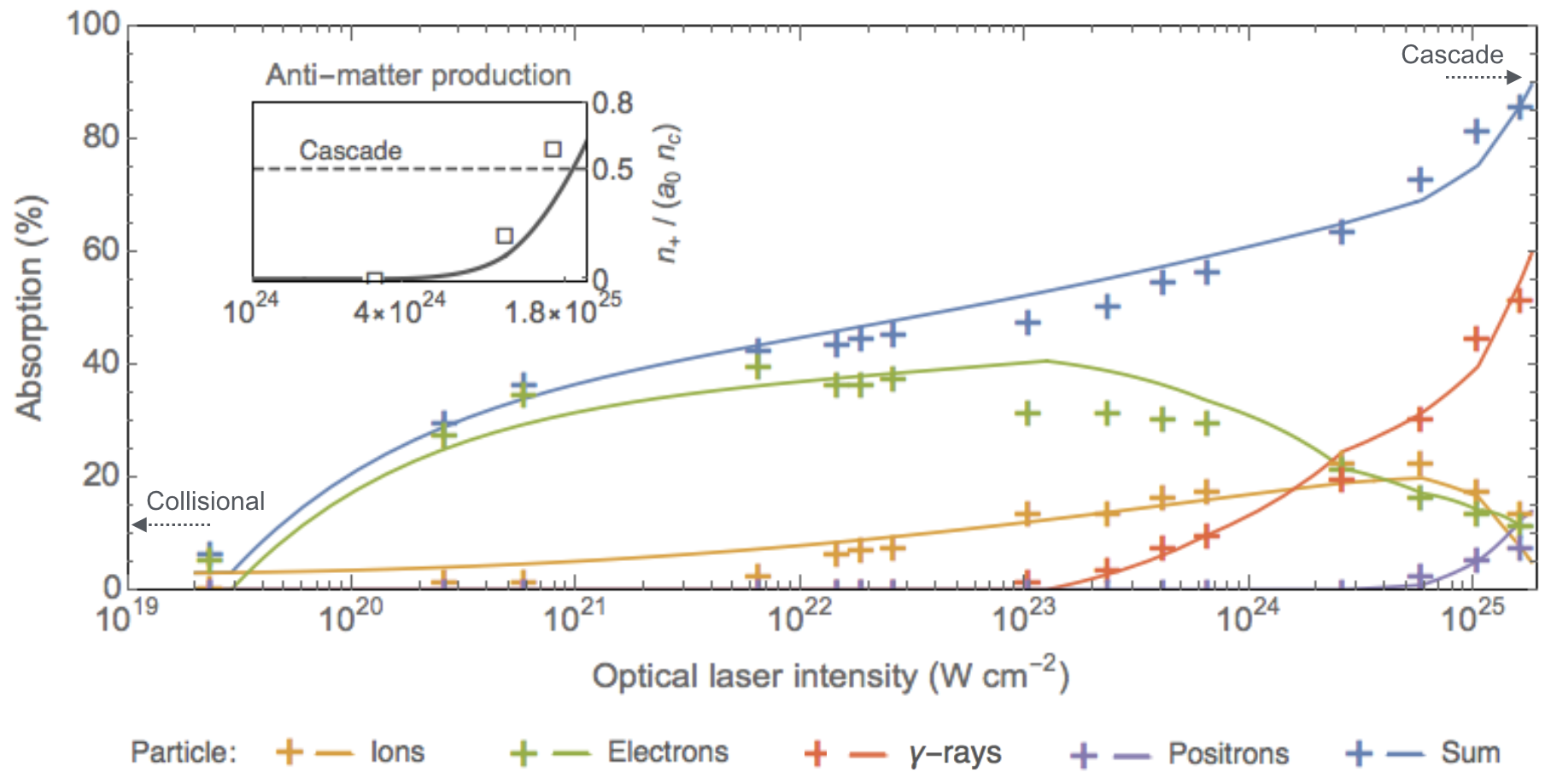}}
	\caption[Absorption across all optical intensity scales]{
		\textbf{Absorption across all optical intensity scales.}
		The quantum conversion efficiency of intense photon flux into energetic
		particles via absorption is presented across six orders-of-magnitude
		in optical laser intensity (solid -- model and markers -- data from
		massive-scale QED-PIC simulations~\cite{Arber2015}). A cascade of
		electron-positron pairs reaching
		$4\times 10^{24}\ \mathrm{positrons}\ \mathrm{cm^{-3}}$ is triggered
		by highly-efficient \gr~production at \casc.
		Mass of the self-created particles accumulates at the interface
		between the laser and quantum electrodynamical plasma to produce
		an inflection in ionic absorption at $5 \times 10^{24}~ \wcm$.
		}
	\label{fig:abs}
	\end{figure*}
Creation of prolific pairs in the absorption region not only provides
new sources of current, but ultimately also screens the supercritical
matter from the incident electromagnetic wave. When the density of
these self-created particles exceeds the relativistically critical density,
the electromagnetic fields responsible for QED processes are damped as,
	\begin{equation}
	a_0 \rightarrow
	 a_0 \sqrt{ a_0 \ncr/(\ne + 2 n_+) } =
	 a_0/\sqrt{1 + 2 P_\pm n_\gamma/ \ne},
	\end{equation} 
where the pair creation probability $P_\pm$ relates the \gr~number density $n_\gamma$
to the positron density $n_+$ according to 
$n_+/\ne = P_\pm n_\gamma / \ne$.  The normalized lepton density effectively grows by
	\begin{equation}
	\nh/\ne \rightarrow
		\nh/\ne 
		\left(
		1 + 2 P_\pm n_\gamma/\ne
		\right).
	\end{equation}
The secondary effect of the increasing electron and positron mass density on
the ions will be described in the following section.

\section{Comparison to QED-PIC Simulation Results}

We use the classical value of $\nh/\ne = 0.3$ to facilitate comparison to the
simulation results.  Since $\ne$ is scaled linearly with $a_0$ to satisfy the
optically-thick condition, $\nh/\ne = \text{const.}$ is consistent with the
well-known $S$-scaling of electron dynamics~\cite{Pukhov2004,Gordienko2005a}. 
This quantity corresponds to the fraction of electrons which escape the bulk
matter and are accelerated by the laser to relativistic energies and is the
only free parameter of the model. Deriving it \emph{ab initio}
is a long-standing topic of interest in the field of high-power laser-plasma physics
which is beyond the scope of the present work. We find the absorption model best
matches the simulation data when this is equal to $0.3$, 
well within the range of values $\simeq 0.1 - 0.5$ typically reported in the
literature~\cite{Kluge2011,Kemp2014,Robinson2014}.

In Fig. \ref{fig:grb} snapshots are shown of the normalized spatial energy-density
of electrons, ions, \gr s, and positrons which illustrate the classical to QED plasma
transition.

Fig. \ref{fig:abs} quantitatively compares predictions of the absorption model
to 15 high-resolution multidimensional QED-PIC simulation~\cite{Arber2015} results.
Extending to the optical intensity in which \gr s are first produced, around
$10^{23} ~\wcm$, the model is seen to accurately describe classical relativistic
laser-matter interactions, reproducing the well-known $S$-scaling of electron
dynamics~\cite{Pukhov2004} while also correctly accounting for ion dynamics. 
In the intensity interval from $10^{23} ~\wcm $ to $3\times 10^{24} ~\wcm$, where
the first positrons are produced (in close agreement with the $\chi\sim 1$ region
shown in Fig. \ref{fig:qed}), the total absorption less that taken by the ions is
constant to within a few percent which confirms the model's picture of energy flow 
as depicted in Fig. \ref{fig:schem}. While both $n_+/\ne$ and $f_+<0.01$, this
intensity scale is within a factor of order-unity of the value first suggested
in \citet{Bell2008}.

Above this intensity, our findings show that the effect of profilic pair creation
is to increase the energy absorbed as \gr s, as the self-created leptons accelerate
and radiatively cool in the same way as electrons originating in the plasma layer.
In addition, the mass of these particles accumulates at the interface between the
laser and QED plasma.  As a consequence of the self-created inertia, the interface
velocity $\betaHB$ slows and an inflection in the ion absorption is produced.
This novel QED plasma dynamic which connects lepton creation to ion acceleration 
emerges naturally from the absorption model due to kinematic coupling of electrons
and ions~\cite{Levy2013PoP,Levy2014} and points to an optimal parameter space for
future QED ion acceleration schemes.
The absorption curves shown in Fig.~\ref{fig:abs} thusly elucidate how optically-thick
matter responds to laser illumination by creation and acceleration of particles
across six orders-of-magnitude in optical intensity.

While integration over the QED rates must in general be carried out numerically,
our results can be closely approximated in terms of $\iOneEight$, the laser's intensity
in units of $10^{18}\ \wcm$, as,
	\begin{align}
		\begin{split}
		f ={}& 9.9\times10^{-5} \iOneEight^{1/2}-\frac{2.6}{\iOneEight^{1/2}}+0.48
		\label{eqn:fEqnsTotFit}
		\end{split}
		\\
		\begin{split}
		f_i ={}& -4.7\times 10^{-12} \iOneEight^{3/2}+7.6\times10^{-5} \iOneEight^{1/2} \\
				& -\frac{0.96}{\iOneEight^{1/2}}+\frac{3.3}{\iOneEight^{0.96}}+0.087
		\label{eqn:fEqnsIFit}
		\end{split}
		\\
 		\begin{split}
 		f_e ={}& 3.1\times 10^{-12} \iOneEight^{3/2}-1.3\times10^{-4} \iOneEight^{1/2} \\
 				&-\frac{1.3}{\iOneEight^{0.34}}+0.45
 		\label{eqn:fEqnsEFit}
 		\end{split}
 		\\
		\begin{split}
		f_\gamma ={}& 1.3\times 10^{-13} \iOneEight^{3/2}+1.4\times10^{-4} \iOneEight^{1/2} \\
					&+\frac{2.3}{\iOneEight^{1.3}}-0.015
		\label{eqn:fEqnsGamFit}
		\end{split}
		\\
		\begin{split}
		f_+ ={}& 2.1\times 10^{-12} \iOneEight^{3/2}-6.1\times 10^{-6} \iOneEight^{1/2} \\
				&+\frac{0.21}{\iOneEight^{1.9}}+1.4\times10^{-3}
		\label{eqn:fEqnsPosFit}
		\end{split}
	\end{align}
where the $R^2$ value associated with each curve is $0.99$.
Processes occurring $>\lp$ downstream from the laser-matter interface are abstracted
from this analysis~\cite{Levy2013PoP,Levy2014}. 
\Cref{eqn:fEqnsTotFit,eqn:fEqnsIFit,eqn:fEqnsEFit,eqn:fEqnsGamFit,eqn:fEqnsPosFit}
for $f, f_k$ therefore provide initial conditions to all current and future modeling
efforts which make use of dense laser-driven particle beams. 

The solid target effectively vanishes behind a screen of self-created particles when
the pair density approaches the initial electron density. To assess this Fig. \ref{fig:abs}
(inset) compares the model-predicted $n_+$ to the positron density near the supercritical
interface measured directly in the simulations. Agreement better than 10\% is observed
all the way to the electron-positron pair cascade at \casc.
In this optical field the positron production achieves $4\times 10^{24}\ \mathrm{cm^{-3}}$,
providing an anti-matter source ~$10^{6}~\times$ denser than of any known photonic
scheme~\cite{Chen2015,Sarri2015}. Furthermore, since the triggering of a pair cascade
is believed to set an ultimate upper limit on attainable electromagnetic field
intensity~\cite{Fedotov2010a,Bulanov2010}, these findings offer further insights into
how the ubiquitous scenario of light absorption/reflection from opaque matter will work
in the most extreme conditions which could ever be achieved on Earth.

\section{Conclusion}

We have shown that by coupling strong-field QED to a classical kinematic theory
of laser-matter interactions, we may predict the  optically-thick absorption to
electrons, ions, \gr s and positrons across all high-power laser intensity scales.
The model, verified by massive-scale QED-PIC simulations, demonstrates novel features
of the plasma response which arise only because of this coupling. At ELI
intensities~\cite{ELI}, the bulk of the laser energy is absorbed to \gr s, leading
to the production of pair plasmaso dense that self-created lepton inertia slows the
velocity of the interface between vacuum and matter. The ion absorption reaches a
maximum which reveals optimal parameter space for ultra-high-field ion accelerator
applications.
These findings thereby lay the groundwork necessary to understand how dense
laser-driven particle beams can be applied to radiotherapy~\cite{Bulanov2002,%
Fiuza2012a,Giulietti2015}
at the petawatt-scale, to the study of nuclear interactions with high density
\gr s~\cite{Ridgers2012,Ledingham2007}, and to scaled laboratory studies of black hole and pulsar
winds~\cite{Blandford1977EM,Bulanov2015}.

	M. C. L. thanks Stephen B. Libby for useful discussions early in the project.
	M. C. L. thanks the Royal Society Newton International Fellowship for support, and the EPSRC Plasma High-End Computing Consortium and University of Oxford Advanced Research Computing (ARC) facility for  computational resources.
	T. G. B. thanks the Knut and Alice Wallenberg Foundation (KAWF) for support.
	N. R. thanks the EPSRC for the support.
	C. P. R. acknowledges support from EPSRC grant number EP/M018156/1.
	A. I. thanks the KAWF and the Olle Engkvist Foundation, grant 2014-744.
	M. M. thanks the KAWF and Swedish Research Council, grants 2012-5644 and 2013-4248.

\appendix
\section{QED-PIC Simulations}

The numerical simulations were carried out using the massively-parallel quantum
electrodynamical particle-in-cell (QED-PIC) code EPOCH~\cite{Ridgers2014,Arber2015}.
EPOCH directly solves the fully-relativistically-correct Lorentz force equation
and full set of Maxwell's equations, thus capturing the relevant kinetic physics
of the classical high-power laser-matter interaction.

In EPOCH QED effects are coupled semi-classically to the PIC workings using a
Monte-Carlo algorithm which describes quantum radiation emission and
electron-positron pair production. The code implements the quasistatic and
weak-field approximations to the QED rates, as described in \cite{Ridgers2014},
which are consistent with the laser-plasma situation. The separation of
energy scales between the \gr~ photons produced by synchrotron emission
and the (optical) photons associated with the high-power laser allows a
treatment of the stochastic processes according to the model of \citet{Baier1968}.

We have carried out 15 multidimensional
simulations using laser and supercritical matter conditions consistent with
the above, over the intensity interval $\sim 10^{19}-10^{25} \ \wcm$.
The code is configured to run in two spatial dimensions for computational
efficiency with a simulation box size of $40 \laml$ in the transverse
$x$ coordinate and $40 \laml$ in the axial $z$ coordinate, each running
from $-20 \laml$ to $20 \laml$, where the laser wavelength is $\laml = 1\um$.
The simulation is assumed to be uniform in the $y$ direction, mimicking a
laser spot with spatial extent in $y$ much greater than $\laml$.

The simulations use a high-power laser pulse whose intensity is varied across
15 values in the range of $\sim10^{19} - 10^{25}\ \wcm$.  The beam is
linearly-polarized in the simulation plane and modeled using a  $8^{th}$-order super-Gaussian 
(\ie square) transverse spatial profile, with spot diameter at focus of $2 r_l = 16\um \gg \laml$.
The temporal profile of the laser is modeled using a $4^{th}$-order super-Gaussian
centered at 50 fs with a full-width half-max of 40 fs. 

The interaction is simulated for 50 $\tau_l$ (optical cycles) which captures
the rise and fall of the laser pulse and confirms that a steady-state of
absorption -- in which $d f_k/d \tau_l \simeq 0$ where $f_k$ is the absorption
into the $k^{th}$ particle type as defined in equation (1) -- is reached.

The supercritical target is modeled as a fully-ionized slab of cold plasma
having electron density $1.01 \sqrt{2} a_0 \ncr$ with ion charge-to-mass
ratio $Z/A=1$.  We have confirmed the matter is opaque to the high-power
laser light, as it should be. The slab is situated within $0 \leq  z/\laml \leq 19$
and $|x/\laml| \leq  19$. The $1\laml$ margin of vacuum is maintained along
each coordinate in order to mitigate any numerical effects related to the
simulation boundaries, which are configured to transmit electromagnetic
radiation and particles.

To facilitate comparison to the model we use the classical $\nh/\ne = 0.3$
and the $\refl$ and $\betaHB$ quantities measured directly in the simulations
in Fig.~\ref{fig:abs}.

The following steps have been taken to ensure that the numerical methods
used in the simulations result in an accurate description of the physics.
The simulation timestep is determined by the Courant condition multiplied by a
factor of 0.5 which enhances stability. The spatial resolution is 
$20$ cells$/\laml$ along both spatial coordinates, corresponding to
3 cells$/\lp$ where the plasma relativistic collisionless skin depth
$\lp  = a_0^{1/2} c/\omega \sqrt{\ncr/\ne}$ (where $c$ is the speed of light,
$\ne$ is the initial plasma electron density, $\ncr = \varepsilon_0 m \omega^2/e^2$
is the critical density, $\omega$ is the laser angular frequency, $m$ is the
electron mass, $\varepsilon_0$ is the vacuum permittivity, and $e$ is the
fundamental charge). 
Therefore all relevant physical scales are resolved. We have confirmed that
the simulation results for all quantities of interest have converged using this
resolution.

Due to the well-known infrared divergence of the photon emission rate,
EPOCH allows the specification of a lower-limit energy for photon particles to
be cast to the simulation grid. This cutoff is taken to be 50 keV, for which
we have confirmed that the absorption values converge.

The code is configured to operate in collisionless mode, since the collisional
mean-free path of relativistic particles in the vacuum absorption region we are
interested in is much longer than the region scale-size. To confirm this, we have
carried out several collisions-on simulations and found no appreciable difference
in the quantities of interest.

Particles are fully kinetic and are represented using 210 electrons per cell and
90 ions per cell, meaning there are $\sim 10^8$ macroparticles in the simulations
at the initial timestep. The higher  mass of the ions means they can be efficiently
modeled using fewer particles, and the particle weighting has been adjusted
to preserve charge neutrality at $t=0$ in the simulation. We have carried out an
extensive survey of conditions and verified that all quantities of interest
converge using this value.

The diagnostic which is used to calculate the absorption in the simulations
tracks the cumulative kinetic energy coupled into each of the $k$ particle species
over time. These values are normalized to the total field energy injected into
the simulation box for an ``empty'' run which is absent the supercritical target.
Following the interaction between the laser and matter, which occurs at around
150~fs due to the hydrodynamic effect, this procedure directly yields a clean
measurement for $f_k$. 
For the steady-state interaction this has been demonstrated in previous works to
be exactly equivalent to equation (1) of the main manuscript text~\cite{Levy2013PoP,Levy2014}.

\mclb{The particle temperatures are calculated using the reciprocal slope of a
fit-line matched to the quasi-exponential portion of the energy distribution at 130fs.}
The hole-boring velocity $\betaHB c$ is calculated by tracking the supercritical
interface at the simulation midplane $x=0$ during the steady-state laser-plasma
interaction from 100-130 fs. The supercritical interface axial location $z_\text{hb}$
is defined by $\sqrt{2} a_0 \ncr/\ne = 1$ so the velocity of the interface is
calculated as $\betaHB c = d z_\text{hb} / dt$. 

%

\end{document}